\documentclass[amsmath, twocolumn, showpacs, aps, prl]{revtex4-1}
\usepackage{bm}
\usepackage{graphics}
\usepackage{graphicx}
\usepackage{color}

\newcommand{\bsigma}{{\boldsymbol \sigma}}
\newcommand{\bnabla}{{\boldsymbol \nabla}}

\newcommand{\brho}{{\boldsymbol \rho}}

\begin{document}

\title{RKKY Interaction On Surfaces of Topological Insulators \\
 With Superconducting Proximity Effect}

\author{Alexander A. Zyuzin and Daniel Loss}

\affiliation{Department of Physics, University of Basel,
Klingelbergstrasse 82, CH-4056 Basel, Switzerland}

\pacs{73.20.-r, 74.25.Ha, 75.10.Hk, 75.50.Lk}

\begin{abstract}
We consider the Ruderman-Kittel-Kasuya-Yosida (RKKY) interaction between magnetic impurities 
on the surface of a three-dimensional topological insulator with proximity induced superconductivity. 
A superconductor placed on the top of the topological insulator induces a gap in the surface electron states and 
gives rise to a long-ranged in-plane antiferromagnetic RKKY interaction. This interaction is frustrated due to strong spin-orbit coupling, 
decays as $1/r$ for $r<\xi$, where $r$ is the distance between two magnetic impurities and $\xi$ the superconducting coherence length, and 
dominates over the ferromagnetic and Dzyaloshinskii-Moriya type interactions for $r>\xi$.
%if the distance between magnetic impurities is larger than superconducting coherence length. 
%inversely proportional to the distance between magnetic impurities if the distance between magnetic impurities is smaller than superconducting coherence length, and dominates over the ferromagnetic and %Dzialoshinskii-Moriya type interactions if the distance between magnetic impurities is larger than superconducting coherence length. 
We find the condition for the Yu-Shiba-Rusinov intragap states that are bound to the magnetic impurities. 
\end{abstract}
\maketitle

{\emph{Introduction}}.~Inducing Cooper pairs in systems possessing strong spin-orbit interaction 
such as topological insulators  \cite{bib:Review1, bib:Pankratov, bib:Kane_Fu} and Rashba spin-orbit coupled films or nanowires 
\cite{bib:AliceaReview} 
%Pankratov, O. A., S. V. Pakhomov, and B. A. Volkov, 1987,
%Solid State Commun. 61, 93.
via proximity effect of a superconductor \cite{bib:Kopnin, bib:Efetov, bib:Buzdin} can lead to topological superconductivity \cite{bib:Review3}.
Topological superconductors host intragap Majorana fermions localized at topological defects
\cite{bib:Volovik-Review}.

\begin{figure}[top]
\centering
     \includegraphics[width=80mm]{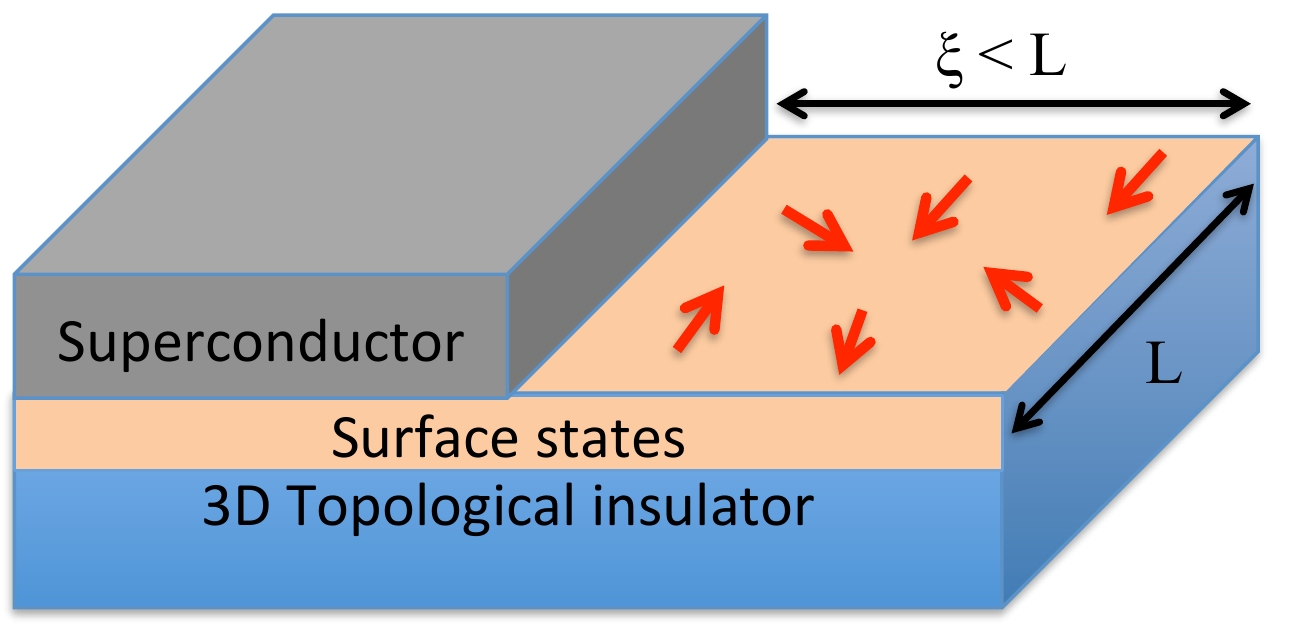}
\caption{An $s$-wave superconductor placed on top of a three-dimensional TI. 
  Proximity induced superconductivity in the surface states of the TI mediates the RKKY interaction between magnetic impurities (red arrows). Here the length of the contact, $L$, can be much larger than the coherence length of the superconducting surface states $\xi$.
}\label{fig:1}
\end{figure}

Motivated by the intense research on the interplay of topological insulators (TI), magnetism, 
%magn:
\cite{bib:Zhang1, bib:Biswas, bib:Garate, bib:Zhang2, bib:Abanin, bib:Efimkin, bib:Burkov, bib:Fan, bib:Melnik}, 
%Burkov
%Burkov, A. A. & Hawthorn, D. G. Spin and charge transport on the surface of a
%topological insulator. Phys. Rev. Lett. 105, 066802 (2010).
%Fan
%Fan, Y. et al. Magnetization switching through giant spin-orbit torque in a
%magnetically doped topological insulator heterostructure. Nature Mater. 13,
%699Ð704 (2014).
%Nature Samarth
%Spin-transfer torque generated by a topological
%insulator
%A. R. Mellnik1, J. S. Lee2, A. Richardella2, J.L.Grab1, P. J. Mintun1, M. H. Fischer1,3, A.Vaezi1, A.Manchon4, E.-A.Kim1, N. Samarth2
%& D. C. Ralph1,5
%
and superconductivity \cite{bib:Review1, bib:Fu, bib:Linder}, 
%sc:
%Kane_Fu
%Fu, L., and C. L. Kane, 2008, Phys. Rev. Lett. 100, 096407.
%hasan Rev. Mod. Phys. 82, 3045 (2010).
%Linder, J., Y. Tanaka, T. Yokoyama, A. Sudbo, and N. Nagaosa,
%2010, Phys. Rev. Lett. 104, 067001.
%
%
we study here the RKKY interaction between magnetic impurities (localized spins) on the surface of a three-dimensional TI with proximity induced superconductivity, see Fig.~\ref{fig:1}.
The effect of  strong spin-orbit interaction at the surface of a TI on the RKKY interaction without superconductivity is well-studied theoretically \cite{bib:Zhang1, bib:Biswas, bib:Garate, bib:Zhang2, bib:Abanin, bib:Efimkin}. 
If the system is at the charge neutrality point, {\it i.e.}, the chemical potential is at the Dirac point, it is  predicted that the localized spins can form a ferromagnetic order with magnetization pointing normal to the surface of the TI, if the out-of-plane anisotropy of the RKKY interaction is stronger than the in-plane one. Out-of-plane ferromagnetic order breaks time-reversal symmetry and opens energy band gaps in the surface states. On the other hand, the in-plane interaction is frustrated. It may result in a spin-glass state \cite{ bib:Galitski, bib:SG} if the in-plane anisotropy is stronger than the out-of plane one \cite{bib:Abanin}.

The effect of magnetic impurities deposited on the surface of a TI on the  band gap and the resulting magnetic order is under experimental debate \cite{bib:Exp1, bib:Exp2, bib:Exp3, bib:Exp4, bib:Exp5, bib:Exp6, bib:Exp7, bib:ExpSTM}.
Alternatively, the locally induced electronic spin density can be measured by NV-center atomic force microscope tips, allowing one to experimentally probe the spin-spin correlations independently of the magnetic impurities
on the surface \cite{bib:Peter}.
The method of spin-polarized scanning tunneling microscope for probing the surface magnetic properties of TI is well
established experimentally \cite{bib:ExpSTM}.

In the present paper, we consider an $s$-wave superconductor placed on the surface of a three-dimensional TI. 
The proximity effect can lead to a gap in the density of surface states of the TI forming a two-dimensional superconductor with strong spin-orbit interaction.
It is well known that in  s-wave superconductors the indirect RKKY coupling between localized spins  has contributions from  Friedel oscillations and from an isotropic antiferromagnetic-type interaction 
\cite{bib:Anderson_Suhl, bib:Abrikosov, bib:Aristov, bib:Galitski}. 

We show that such antiferromagnetic interactions between localized spins in TIs with proximity induced superconductivity are frustrated due to strong spin-orbit interaction and  the RKKY energy is minimized
if the localized spins point in the plane of the superconducting surface normal to the line connecting them.

We find that superconductivity strongly increases the antiferromagnetic interaction and makes it dominant over the ferromagnetic and Dzyaloshinskii - Moryia types of ordering, if the separation between two magnetic impurities is larger than the superconducting coherence length $\xi$ of the surface states. The interaction between localized spins is exponentially suppressed in this case. 
However, it was shown that antiferromagnetic interaction in $s$-wave superconductors gets  enhanced due to the formation of Yu-Shiba-Rusinov bound states \cite{bib:Glazman},
already at distances between localized spins smaller than the coherence length. We expect similar results for TIs.

{\emph{Description of the model}}.~The low-energy behaviour of ballistic electrons at the surface of 
a three-dimensional TI with proximity-induced superconducting gap can be described by the BCS Hamiltonian
\begin{equation}\label{Ham1}
H = \frac{1}{2}\int d^2 { r}   \Psi^{\dag}(\mathbf{ r}) \bigg\{ (-iv_F \bsigma \times \hat{\mathbf{ z}}\cdot\bnabla_\mathbf{ r} -\mu)\tau^z +\Delta\tau^{x}\bigg\}\Psi(\mathbf{ r}),~
\end{equation}
where $\Psi^{\dag}(\mathbf{ r}) = (\psi^{\dag}_{\uparrow}(\mathbf{ r}),\psi^{\dag}_{\downarrow}(\mathbf{ r}), \psi_{\downarrow}(\mathbf{ r}),  -\psi_{\uparrow}(\mathbf{ r}))$ is the Nambu
operator of an electron at the superconducting TI surface; $\sigma^a$ and $\tau^a$ ($a=x,y,z$) are the Pauli matrices acting on the spin and the particle-hole degrees
of freedom, respectively; $\hat{\mathbf{ z}}$ is a
unit vector in the direction perpendicular to the surface; $v_F$, $\mu$, and $\Delta$ are the Fermi velocity, chemical potential, and proximity induced superconducting gap of the surface states, respectively. We set $\hbar =1$.
Generally, $\Delta$ depends on the properties of  the
superconductor-TI interface \cite{bib:Kopnin, bib:Buzdin, bib:Efetov}. Here we will not be interested in the exact expression of the gap and assume it to be a coordinate-independent  positive constant.

The local anisotropic exchange interaction between magnetic impurities with classical spin $S$, localized on the proximitized TI surface, and the  spin density  of the surface electrons of the TI is described by the Hamiltonian
\begin{eqnarray}\label{Ham3}\nonumber
H_{\mathrm{ex}} &=& \frac{1}{4}\int d^2{ r} 
\Psi^{\dag} (\mathbf{ r})\sum_{j} [J_zS_{jz}\sigma^z\\
&+&J_{x}(S_{jx}\sigma^x+S_{jy}\sigma^y) ]\delta(\mathbf{ r}-\mathbf{ r}_j) \Psi(\mathbf{ r}).
\end{eqnarray}
The direction of the localized spin at  point $\mathbf{ r}_j$ is defined by the vector $\mathbf{ S}_j = (S_{jx}, S_{jy}, S_{jz})$.  
The exchange  is assumed to be anisotropic, with the in- and out-of plane coupling constants given by $J_{x}=J_y$ and $J_z$, respectively. 

From Eqs.~(\ref{Ham1}) and (\ref{Ham3}) we see that the particle-hole symmetry is preserved in 
the presence of  exchange interactions between the impurity and  electron spins. 
The interaction of the electron spin with the in-plane component of the localized spin locally breaks the in-plane rotational symmetry, while
the interaction with the out-of-plane component locally breaks the time-reversal symmetry.

It is known that magnetic impurities can give rise to intragap Yu-Shiba-Rusinov bound states in the superconductor \cite{bib:Yu, bib:Shiba, bib:Rusinov, bib:Sakurai, bib:Bal-review}.
Changing the exchange interactions one  
can tune these states through  zero energy, counted from the position of the Fermi level, leading to a  quantum phase transition. The spin quantum number of the ground state changes \cite{bib:Balatsky, bib:Bal-review}. Magnetic impurities in TIs with proximity-induced superconductivity can support such bound states as well. The 
intragap states at the surface of the three-dimensional TI with superconducting proximity effect were studied numerically \cite{bib:TSCRKKY}. Here we will give general analytical conditions for the  bound state at the middle of the gap.

The energy levels bound to the localized magnetic impurities are defined by the poles of the electron Green function with the 
self-energy that takes into account multiple scatterings off a magnetic impurity,
\begin{equation}\label{Resonance}
\mathrm{Det}\bigg[ 1- \frac{1}{2}(J_zS_{jz}\sigma^z+ J_{x}(S_{jx}\sigma^x+S_{jy}\sigma^y))G(\omega_n, \mathbf{ r}=0)\bigg]=0,~
\end{equation}
where $\mathrm{Det}$ means determinant.
It is convenient to write the Green function of the TI electrons in Nambu space as
$G(\omega_n, \mathbf{ r})=\int \frac{d\mathbf{ p}}{(2\pi)^2} G(\omega_n, \mathbf{ p}) e^{i\mathbf{ p}\cdot\mathbf{ r}}$, where in the momentum representation one has
\begin{equation}\label{Green}
 G(\omega_n, \mathbf{ p})=-\frac{1}{2}\sum_{\lambda =\pm 1}\left(1+\lambda  \bsigma \times \hat{\mathbf{ z}}\cdot\hat{\mathbf{p}}\right)\frac{i\omega_n+\xi_{\lambda}\tau^z+\Delta\tau^x}
 {\omega^2_n+\xi_{\lambda}^2+\Delta^2}.~
\end{equation}
Here, summation is performed over two bands with opposite helicities $\lambda=\pm 1$, $\hat{\mathbf{ p}}=\mathbf{ p}/|\mathbf{ p}|$ is the unit vector in the direction of the in-plane momentum 
of electron, $\omega_n=\pi T(2n+1)$ is the fermionic Matsubara frequency, $T$ is the temperature, and $\xi_{\lambda}=\lambda v_F|\mathbf{ p}|-\mu$. 
The Green function at point $\mathbf{ r}=0$ 
becomes
\begin{widetext}
\begin{eqnarray}\label{GreenR}
G(\omega_n,\mathbf{ r}=0)=
\frac{\sqrt{\omega_n^2+\Delta^2}}{2\pi v_F^2}\bigg[\tau^z-\mu\frac{i\omega_n+\Delta\tau^x}{\omega_n^2+\Delta^2}\bigg]\mathrm{arctg}\frac{\mu}{\sqrt{\omega_n^2+\Delta^2}}
-\frac{\mu}{2\pi v_F^2}\bigg[\tau^z+\frac{i\omega_n+\Delta \tau^x}{\mu}\bigg] \ln \frac{\Lambda}{\sqrt{\omega_n^2+\Delta^2+\mu^2}}. ~
\end{eqnarray}
\end{widetext}
The high energy cut-off is defined by the surface states band width, $\Lambda$. Two quasiparticle energy levels show up inside the gap symmetrically below and above the Fermi level as the strength of
$J_{x,z}$
increases. 
We substitute expression (\ref{GreenR}) into (\ref{Resonance}), and perform the analytical continuation $i\omega_n \rightarrow \omega +i\delta$. We find the necessary condition for the zero 
energy, $\omega =0$, Yu-Shiba-Rusinov bound state in the superconducting surface of the TI,
\begin{eqnarray}\label{E3}
&(J_z^2-J_{x}^2)S^2_{jz}+J_{x}^2S^2& \\\nonumber
&=\frac{(4\pi v_F^2)^2}{\Delta^2+\mu^2}\bigg[\mathrm{arctg}^2(\mu/\Delta)+\ln^2(\frac{\Lambda}{\sqrt{\Delta^2+\mu^2}})\bigg ]^{-1}.&
\end{eqnarray} 
At $\mu =0$ the dependence of the exchange interactions  on the proximity gap $\Delta$ is given by
\begin{equation}\label{E1}
\left[(J_z^2-J_{x}^2)S^2_{jz}+J_{x}^2S^2\right]^{1/2} = \frac{4\pi v_F^2}{\Delta \ln (\Lambda/\Delta)}~.
\end{equation}
The interaction constant is inversely proportional to the superconducting gap
up to the logarithmic prefactor.
In the limit when the chemical potential is much larger than the gap but smaller than the band width, $\Lambda>\mu>\Delta$, we obtain, $\left[(J_z^2-J_{x}^2)S^2_{jz}+J_{x}^2S^2\right]^{1/2} \simeq \frac{4\pi v_F^2}{\mu \ln (\Lambda/\mu)}[1+\pi \Delta/2\mu\ln^2(\Lambda/\mu)]$. An increase of $\mu$
decreases the values of $J_{x,z}$ at which the  bound states are pinned to the middle of the superconducting gap. 
To compare, the condition for the bound state at the Fermi level in an $s$-wave superconductor is given by
$JS\pi \nu/2 =1$, where $\nu$ is the normal-state density of states at the Fermi level and the exchange coupling $J$ is isotropic \cite{bib:Yu, bib:Shiba, bib:Rusinov, bib:Sakurai}.

{\emph{RKKY interaction}}.~Far from the resonance Eq.~(\ref{E3}), 
the RKKY interaction between two localized spins $\mathbf{ S}_1$ and $\mathbf{ S}_2$  at points $\mathbf{ r}_1$ and $\mathbf{ r}_2$, respectively, and separated by the distance $r=|\mathbf{ r}| \equiv |\mathbf{ r}_1-\mathbf{ r}_2|$  is defined as \cite{bib:Abrikosov}
\begin{equation}
\mathcal{E}(\mathbf{ r}) = \frac{T}{2}\sum_{n;a,b} \mathrm{Tr} \bigg\{J_{a}S_{1a}\frac{\sigma^{a}}{2} G(\omega_n,\mathbf{ r})J_{b}S_{2b}\frac{\sigma^{b}}{2}G(\omega_n,-\mathbf{ r})\bigg\},
\end{equation}
where $\mathrm{Tr}$ means the trace over the spin and particle-hole degrees of freedom of the TI electrons.
It is convenient to consider the limit of large chemical potential, $\mu/\Delta \gg 1$, $\mu r/v_F \gg 1$, and the limit 
$\mu=0$ separately. 

{\emph{RKKY interaction for zero chemical potential}}.~At $\mu =0$ the single-electron Green function reads
\begin{eqnarray}
G(\omega_n, \mathbf{ r}) &=& \frac{i\omega_n+\Delta\tau^x}{2\pi v_F^2}K_0\bigg(\frac{r}{v_F}\sqrt{\omega_n^2+\Delta^2}\bigg)\\\nonumber
&+&i\hat{\brho} \cdot\bsigma \tau^z
\frac{\sqrt{\omega_n^2+\Delta^2}}{2\pi v_F^2}K_1\left(\frac{r}{v_F}\sqrt{\omega_n^2+\Delta^2}\right),
\end{eqnarray}
where $\hat{\brho} = \hat{\mathbf{ z}}\times \hat{\mathbf{ r}}$ is an in-plane unit vector that is perpendicular to the unit vector $\hat{\mathbf{ r}}$  pointing along the line connecting  two localized spins, and $K_n(x)$  is the Macdonald function.
We are interested in the zero temperature limit, $T=0$, and transform the sum over frequencies into an integral, $T\sum_n \rightarrow \int \frac{d\omega}{2\pi}$. 
We obtain the RKKY interaction between two magnetic impurities in the form
\begin{eqnarray}\nonumber\label{exchange}
\mathcal{E}(\mathbf{ r}) &=& -\frac{A_{+}(2r/\xi)}{64\pi v_F r^3}\bigg[J^2_{z}S_{1z}S_{2z} + J^2_{x}(\mathbf{ S}_{1}\cdot\hat{\mathbf{ r}})(\mathbf{ S}_{2}\cdot\hat{\mathbf{ r}})\bigg] \\ & +&
\frac{A_{-}(2r/\xi)}{64\pi v_F r^3}J^2_{x}(\mathbf{ S}_{1}\cdot\hat{\mathbf{ \brho}})(\mathbf{ S}_{2}\cdot\hat{\mathbf{ \brho}}),
\end{eqnarray}
where $\xi=v_F/\Delta$ is the superconducting coherence length. The interaction coefficients $A_{\pm}(\alpha)$ are defined through the dimensionless integrals:
\begin{eqnarray}\label{A}\nonumber
A_{\pm}(\alpha) &=&\int_{0}^{\infty}\frac{dx}{\pi^2}\bigg[(x^2+\alpha^2)K_1^2\bigg(\sqrt{x^2+\alpha^2}/2\bigg)\\
&\pm&(x^2-\alpha^2)K_0^2\bigg(\sqrt{x^2+\alpha^2}/2\bigg)\bigg].
\end{eqnarray}

\begin{figure}[t]
\includegraphics[width=8cm] {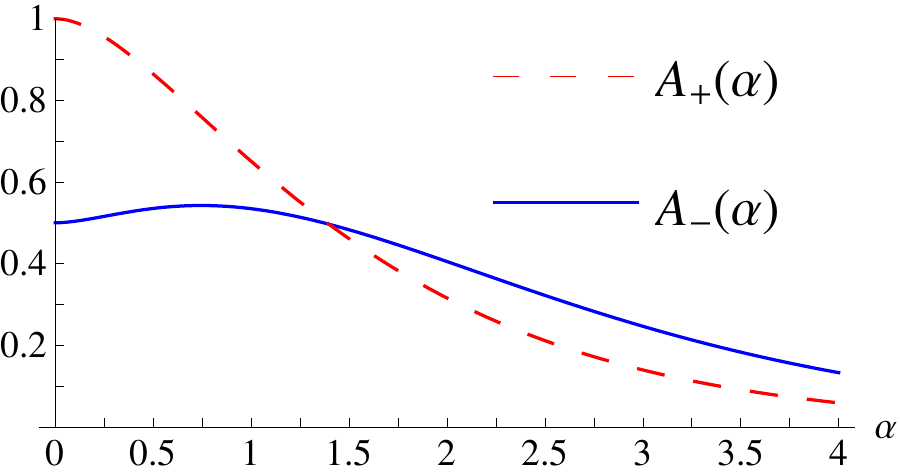}
\caption{The coordinate dependence of the coefficients $A_{\pm}(\alpha)$ at $T=0$ when the surface states of the TI are at the Dirac point, $\mu=0$. 
Here, $\alpha =2r/\xi$, where $\xi=v_F/\Delta$ is the superconducting coherence length. }
\label{fig:2}
\end{figure}
Analytical expressions for the RKKY interaction can be obtained in several limiting cases.
For $\Delta=0$ we recover the known RKKY interaction on the surface of a three-dimensional TI: $A_{+}(\alpha)=1$ and $A_{-}(\alpha)=1/2$ \cite{bib:Zhang1, bib:Biswas, bib:Garate, bib:Zhang2, bib:Abanin}. In this limit the RKKY interaction scales as $1/r^3$. We note that in the limit of small separation $r\rightarrow 0$ the RKKY interaction is cut by the interatomic distance.

For $\Delta \neq0$, we find the asymptotics
$A_{\pm}(\alpha)  \simeq (1\mp \alpha^2)2^{(\pm1-1)/2}$,
if $r<\xi$, {\it i.e.}, $\alpha < 1$, while in the opposite case, $\alpha > 1$, 
we obtain
\begin{eqnarray}\nonumber
A_{+}(\alpha) & \simeq& \sqrt{2\alpha/\pi}e^{-\alpha},\\
A_{-}(\alpha) & \simeq& \alpha A_{+}(\alpha).
\end{eqnarray}
 In Fig. (\ref{fig:2}), we plot the exact numerical solution of the interaction coefficients $A_{\pm}(\alpha)$ as function of $\alpha=2r/\xi$.

Since both coefficients $A_{\pm}(r)$ are positive, the first term on the rhs of Eq.~(\ref{exchange}) favours ferromagnetic ordering between two magnetic impurities with spins pointing normal to the plane,
while the second term favours in-plane ferromagnetic ordering between two magnetic impurities with spins pointing along the line joining them. 
The third term describes the in-plane antiferromagnetic ordering in perpendicular direction to the line connecting the two magnetic impurities \cite{bib:Zhang1, bib:Biswas, bib:Garate, bib:Zhang2, bib:Abanin}.

For $\Delta \neq 0$ the RKKY interaction is exponentially suppressed if the the separation between two magnetic impurities is larger than the superconducting coherence length.
We note that in the presence of the superconducting gap the strength of the in-plane antiferromagnetic coupling between two localized spins increases compared to both in-plane and out-of-plane ferromagnetic couplings, $A_{-}(\alpha)/A_{+}(\alpha) = \alpha$ at $\alpha> 1$.

It was argued that the in-plane frustration might lead to the spin glass state if the in-plane exchange interaction exceeds the out-of-plane one \cite{bib:Abanin}. The quantum critical point corresponding to the transition to the glassy state was estimated as $J_x/J_z \approx 1.3$ at $T=0$. For $\Delta \neq 0$, however, we expect the spin glass state at lower values of $J_x/J_z\lesssim 1$, for an average distance between localized spins on the order of $\xi$.

{\emph{RKKY interaction for finite chemical potential}}.~Let us now turn to the large chemical potential limit defined by $\mu \gg \Delta$ and $\mu r/v_F\gg 1$. In this case the Green function in 
Eq.~ (\ref{Green}) 
can be projected onto the band with positive helicity, $\lambda=1$,
\begin{eqnarray}\nonumber
G(\omega_n, \mathbf{ r})&=&\sqrt{\frac{\mu}{8\pi v_F^3r}} e^{-\frac{r}{v_F}\sqrt{\omega_n^2+\Delta^2}
  +i\hat{\brho}\cdot\bsigma \left(r\mu/v_F -\pi/4\right)}\\
&\times&\bigg(\frac{i\omega_n+\Delta \tau^x}{\sqrt{\omega_n^2+\Delta^2}}+i\hat{\brho}\cdot\bsigma\tau^z\bigg),
\end{eqnarray}
where we used $(\hat{\brho}\cdot\bsigma)^2=1$.
Then, the RKKY interaction becomes,
\begin{eqnarray}\label{B}\nonumber
\mathcal{E}(\mathbf{ r})&=& \frac{\mu A_1(2r/\xi)}{(4\pi v_F r)^2}\bigg\{
J_xJ_z\hat{ \brho}\cdot \mathbf{ S}_1\times \mathbf{ S}_2 \cos\left(2r\mu/v_F\right)\\\nonumber&-&
\left[J^2_{z}S_{1z}S_{2z} + J^2_{x}(\mathbf{ S}_{1}\cdot\hat{\mathbf{ r}})(\mathbf{ S}_{2}\cdot\hat{\mathbf{ r}})\right]\sin\left(2r\mu/v_F\right)\bigg\}
\\
&+& \frac{\mu A_2(2r/\xi)}{(4\pi v_F r)^2} J_x^2 (\mathbf{S}_{1}\cdot\hat{\brho})( \mathbf{S}_{2}\cdot\hat{\brho}),
\end{eqnarray}
where the interaction coefficients at $T=0$ are given by
\begin{eqnarray}\nonumber
A_1(\alpha) &=&\alpha K_1(\alpha)-A_2(\alpha),\\
A_2(\alpha) &=& \int_{0}^{\infty}\frac{\alpha^2dx}{x^2+\alpha^2}e^{-\sqrt{x^2+\alpha^2}}.
\end{eqnarray}
\begin{figure}[t]
\includegraphics[width=8cm] {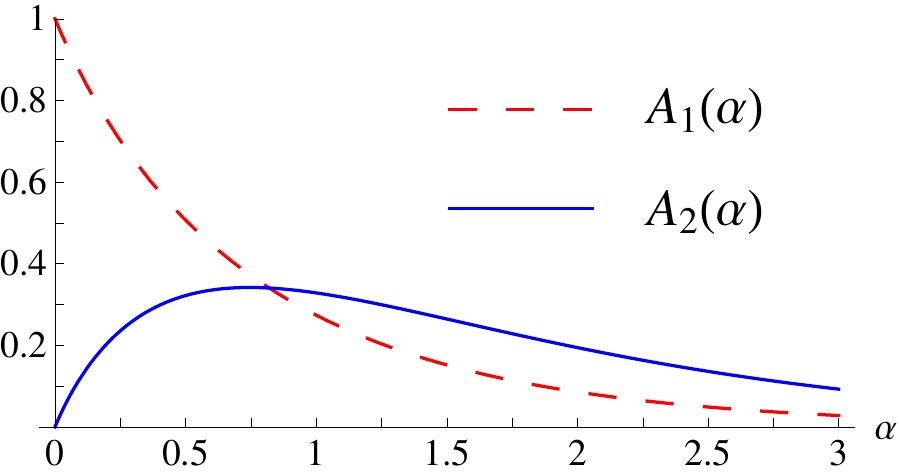}
\caption{The coordinate dependence of  $A_{1,2}(\alpha)$ for $\alpha=2r/\xi$, at $T=0$, and in the 
regime $\mu \gg v_F/r, \Delta$, 
with $\xi=v_F/\Delta$  the superconducting coherence length. 
}
\label{fig:3}
\end{figure}
We note that the RKKY interaction for $r\rightarrow 0$ 
is cut by the Fermi wave length $2\pi/k_F$. In the normal state, $\Delta=0$, one obtains the asymptotic form of the RKKY interaction on the surface of a three-dimensional TI \cite{bib:Zhang1, bib:Biswas, bib:Garate, bib:Zhang2, bib:Abanin}:
$A_1(\alpha)= 1$ and $A_2(\alpha)=0$. In leading order in $v_F/(\mu r)\ll 1$, the finite chemical potential suppresses the antiferromagnetic in-plane coupling, the third term on the rhs of Eq.~(\ref{exchange}), and
gives rise to a Dzyaloshinskii-Moriya type coupling, the third term in Eq.~(\ref{B}). Such exchange anisotropy is similar to the one induced by spin orbit interaction in normal semiconductors \cite{bib:Imamura} and in nanotubes and nanoribbons \cite{bib:Klinovaja}.
For $\Delta \neq 0$ and $\alpha < 1$ the interaction coefficients have the following asymptotic forms,
$A_1(\alpha) \simeq 1-\pi \alpha/2$ and $ A_2(\alpha) \simeq \pi \alpha/2$, while in the opposite limit, $\alpha >1$, they are given by
\begin{eqnarray}\nonumber
A_1(\alpha) &\simeq& \sqrt{\pi/2\alpha}\, e^{-\alpha},\\
A_2(\alpha) &\simeq& \alpha A_1(\alpha).
\end{eqnarray}
In Fig. (\ref{fig:3}), we plot  $A_{\pm}(\alpha)$ as function of $\alpha=2r/\xi$ obtained by exact numerical evaluation.

We see that the superconducting gap at finite chemical potential gives rise to the antiferromagnetic RKKY interaction term (\ref{B}), which in the
limit $\Delta=0$ is smaller than the terms in Eq.~(\ref{exchange}) as $v_F/(\mu r) \ll 1$ \cite{bib:Garate}. This term 
is long ranged: It decays as $1/r$, if $r<\xi$, compared to the usual oscillating terms, which decay as $1/r^2$. 
Similarly to the case of zero chemical potential, superconductivity increases the in-plane antiferromagnetic ordering between two magnetic impurities with spins pointing in the perpendicular direction to the line connecting them.

{\emph{Conclusions}}.~ 
We studied the RKKY interaction on the surface of the three-dimensional TI with proximity-induced superconductivity. 
We showed that in addition to the conventional interaction terms which oscillate at finite chemical potential as $\sin(2\mu v/r)$ and decay as $1/r^2$, superconductivity 
gives rise to an antiferromagnetic long range term, which decays as $1/r$, provided $r<\xi$. 
This in-plane antiferromagnetic ordering between two magnetic impurities with 
spins pointing normal to the line connecting them dominates the RKKY interaction when the distance between the impurities exceeds $\xi$. We find the condition for the
Yu-Shiba-Rusinov bound states in the superconducting surface of the TI.

To conclude, we propose the frustrated antiferromagnetic RKKY interaction as a direct manifestation of the intrinsic properties of a TI with proximity-induced superconductivity 
which gives rise  to a rich
behaviour of the interaction as a function of doping, exchange interaction, and distance and orientation between impurities.
It would be interesting to study the effect of disorder on the RKKY interaction in the TI \cite{bib:Papa, bib:Chesi}.

{\emph{Acknowledgements}}.~We thank L. I. Glazman and A. Yu. Zyuzin for helpful discussions and acknowledge support from the Swiss NF and NCCR QSIT.

\bibliography{references}
\end{document}